\documentclass[a4paper]{jpconf-kb}
\usepackage{amsmath}
\usepackage{iopams}
\usepackage{natbib}
\usepackage{graphicx}

\usepackage{imakeidx}
\makeindex[name=subject,title=Subject index,columns=1]
\makeindex[name=authors,title=Author index,columns=1]

\begin{document}
\graphicspath{{FIGURES/}}

\title{Topological stirring of two-dimensional atomic Bose-Einstein condensates}

\author{A.~C.\ White$^{1,2}$, N.~P.\ Proukakis$^{1}$ \& C.~F.\ Barenghi$^1$}

\address{$^1$ Joint Quantum Centre (JQC), Durham-Newcastle, School of Mathematics and \newline
\phantom{$^1$}Statistics, Newcastle University, Newcastle upon Tyne, NE1 7RU, United Kingdom.}
\address{$^2$ School of Mathematical Sciences, University of Nottingham, University Park, \newline
\phantom{$^2$ }Nottingham, NG7 2RD, United Kingdom.}

\ead{ang.c.white@gmail.com}

\index[authors]{White, A.~C.} \index[authors]{Proukakis, N.~P.} \index[authors]{Barenghi, C.~F.}

\begin{abstract}
We stir vortices\index[subject]{vortices} into a trapped quasi two-dimensional atomic Bose-Einstein condensate\index[subject]{Bose-Einstein condensate} by moving three laser stirrers. We apply stirring protocols introduced by \cite{Boyland2000}\index[authors]{Boyland, P.~L.}\index[authors]{Aref, H.}\index[authors]{Stremler, M.~A.}, that efficiently build in topological chaos \index[subject]{chaos} in classical fluids and are classified as Pseudo-Anosov stirring protocols. These are compared to their inefficient mixing counterparts, finite-order\index[subject]{finite-order}
 stirring protocols. We investigate if inefficient stirring protocols result in a more clustered\index[subject]{clustering}
 distribution of vortices\index[subject]{vortices}. The efficiency with which vortices\index[subject]{vortices}
 are `mixed' or distributed in a condensate is important for investigating dynamics of continuously forced quantum turbulence\index[subject]{turbulence} and the existence of the inverse cascade\index[subject]{inverse cascade}
 in turbulent\index[subject]{turbulence}
 two-dimensional superfluids. 
\end{abstract}

\section{Introduction}

Jupiter's Great Red Spot is a beautiful demonstration of a large spot of vorticity, one of the defining features of two dimensional (2D) classical turbulence\index[subject]{turbulence}, arising from the inverse cascade\index[subject]{inverse cascade} process (\cite{Sommeria1988,Marcus1988}\index[authors]{Marcus,P.~S.}\index[authors]{Sommeria, J.}\index[authors]{Meyers, S. D.}\index[authors]{Swinney, H. L.}). This process describes the flow of energy from small scales, given by the scale at which energy is injected into the system, to larger scales, manifesting in the clustering\index[subject]{clustering} of vortices\index[subject]{vortices} of like-winding. The concept of an inverse energy cascade process was introduced by \cite{Onsager1949}\index[authors]{Onsager, L.}, who showed that clusters of like-signed point vortices\index[subject]{vortices} in a large Onsager gas of many point-vortices\index[subject]{vortices} possess negative temperature. Quantum vortices\index[subject]{vortices} in superfluid He can be realistically modelled as point vortices\index[subject]{vortices} or vortex lines, due their small vortex core size ($\sim1$nm), quantised vorticity and large ratio of inter-vortex spacing to vortex core radius ($\sim10^{5}-10^{6}$).  At large scales greater than the average inter-vortex spacing, three dimensional turbulent quantum fluids are also known to exhibit the same classical Kolmogorov energy spectrum with scaling $k^{-5/3}$ (where $k$ is the wavenumber) as classical fluids  (see \cite{Nore1997,Maurer1998,Stalp1999,Araki2002,Kobayashi2005,Salort2010PF,Baggaley2011PRE}\index[authors]{Nore, C.}\index[authors]{Abid, M.}\index[authors]{Brachet, M.~E.}\index[authors]{Maurer, J.}\index[authors]{Tabeling, P.}\index[authors]{Stalp, S. R.}\index[authors]{Skrbek, L.}\index[authors]{Donnelly, R. J.}\index[authors]{Araki, T.}\index[authors]{Tsubota, M.}\index[authors]{Nemirovskii, S. K.}\index[authors]{Kobayashi, M.}\index[authors]{Tsubota, M.}\index[authors]{Salort, J.}\index[authors]{Baudet, C.}\index[authors]{Castaing, B.}\index[authors]{Chabaud, B.}\index[authors]{Daviaud, F.}\index[authors]{Dedelot, T.}\index[authors]{Diribarne, P.}\index[authors]{Dubrulle, B.}\index[authors]{Gagne, Y.}\index[authors]{Gauthier, F.}\index[authors]{Girard, A.}\index[authors]{H{\'e}bral, B.}\index[authors]{Rousset, B.}\index[authors]{Thibault, P.}\index[authors]{Roche, P.-E.}\index[authors]{Baggaley, A.~W.}\index[authors]{Barenghi, C.~F.} and \cite{Baggaley2012PRL}\index[authors]{Baggaley, A.~W.}\index[authors]{Laurie, J.}\index[authors]{Barenghi, C.~F.}).  For these reasons, one might expect the inverse cascade\index[subject]{inverse cascade} process to occur in other two dimensional quantum fluids, such as Bose-Einstein condensates (BECs)\index[subject]{Bose-Einstein condensate}, where vortices\index[subject]{vortices} also have quantised circulation.  While at large scales turbulent quantum fluids are known to have similarities to turbulent classical fluids for three dimensional systems, for two dimensional systems such a crossover is yet to be established. There have been some numerical simulations working towards this goal in trapped and homogeneous condensates, however, to date  there is no conclusive demonstration of the inverse cascade\index[subject]{inverse cascade} process for harmonically trapped atomic BECs \index[subject]{Bose-Einstein condensate} (\cite{Parker2005,Horng2009,Numasato2010,Numasato2010JLTP,White2012,Bradley2012,Reeves2012,Reeves2012arxiv}\index[authors]{Parker,N~G.}\index[authors]{Adams, C.~S.}\index[authors]{Horng, T.-L.}\index[authors]{Hsueh, C.-H.}\index[authors]{Su, S.-W.}\index[authors]{Kao, Y.-M.}\index[authors]{Gou, S.-C.}\index[authors]{Numasato, R.}\index[authors]{Tsubota, M.}\index[authors]{Numasato, R.}\index[authors]{Tsubota, M.}\index[authors]{L'vov, V. S.}\index[authors]{White, A.~C.}\index[authors]{Barenghi, C.~F.}\index[authors]{Proukakis, N.~P.}\index[authors]{Bradley, A.~S.}\index[authors]{Anderson, B.~P.}\index[authors]{Reeves, M.~T.}\index[authors]{Anderson, B.~P.}\index[authors]{Bradley, A.~S.}).  Unlike in superfluid He, atomic BECs\index[subject]{Bose-Einstein condensate} of low dimensions can be routinely created and manipulated experimentally and so they provide a promising system for such investigations. One important component of any two dimensional turbulent set-up is the introduction of vortices\index[subject]{vortices} into the condensate in a controlled manner, such that their distribution is well-mixed. Finding such a well-mixed vortex distribution will be the focus of this work. 

Recently \cite{White2012}\index[authors]{White, A.~C.}\index[authors]{Barenghi, C.~F.}\index[authors]{Proukakis, N.~P.} showed that the trajectory a single laser stirrer takes through the condensate determines if the resulting vortices\index[subject]{vortices} are clustered or more randomly distributed. We found that a single laser moving on a circular trajectory creates an initial cluster of like-signed vortices\index[subject]{vortices}. While the extent of clustering\index[subject]{clustering} was found to increase with stirring, it did not persist and decayed after the laser stirrer was turned off. This clustering\index[subject]{clustering} of vortices\index[subject]{vortices} manifests due to the trajectory of the laser stirrer, and not as a result of an inverse cascade\index[subject]{inverse cascade} process.  Indeed it was shown that altering the trajectory of the laser stirrer resulted in less-extensive clustering\index[subject]{clustering} and vortices\index[subject]{vortices} could even be created in more randomly distributed configurations.  In this work we employ three laser stirrers and measure the distribution and clustering\index[subject]{clustering} of vortices\index[subject]{vortices} by applying statistical measures of clustering\index[subject]{clustering} previously developed in \cite{White2012}\index[authors]{White, A.~C.}\index[authors]{Barenghi, C.~F.}\index[authors]{Proukakis, N.~P.}.  In particular, we stir the condensate in a manner that is known to efficiently mix classical fluids, known as pseudo-Anosov\index[subject]{pseudo-Anosov} stirring protocols. These are contrasted to finite-order\index[subject]{finite-order} stirring protocols which have been shown to be inefficient mixers for classical fluids.  

In this paper we proceed in the following way. Firstly we give an overview of topological stirring protocols and then outline the model applied to simulate vortex dynamics in atomic BECs\index[subject]{Bose-Einstein condensate}. Finally we introduce statistical measures of clustering\index[subject]{clustering} that are applied to give insight into how two quite different stirring protocols influence vortex dynamics in a surprisingly similar way.  While this is unexpected from the perspective of classical fluids, it is a remarkable demonstration of the potential flow of quantum fluids.   
 
\section{Topological stirring protocols} 
Topological concepts were introduced to address the question of how to most efficiently mix classical fluids.  \cite{Boyland2000}\index[authors]{Boyland, P.~L.}\index[authors]{Aref, H.}\index[authors]{Stremler, M.~A.} determined that for flows with the topology of certain braids, given only the topology of the flow, it is possible to deduce the material stretch rate.  For stokes flow (characterised by Reynolds number, $Re\ll1$), chaotic advection of fluid particles is built into the flow faster by stirring schemes that build in topological chaos\index[subject]{chaos}.   Such protocols are classified as pseudo-Anosov\index[subject]{pseudo-Anosov} stirring protocols and it is established that stirring a fluid in such a way results in exponential stretching of material lines.  Stirring protocols that do not have the correct topology to mix effectively are known as finite-order\index[subject]{finite-order} stirring protocols (\cite{Boyland2000,Finn2003, Gouillart2006}\index[authors]{Boyland, P.~L.}\index[authors]{Aref, H.}\index[authors]{Stremler, M.~A.}\index[authors]{Finn, M.~D.}\index[authors]{Cox, S.~M.}\index[authors]{Byrne, H.~M.}\index[authors]{Gouillart, E.}\index[authors]{Thiffeault, J-L.}\index[authors]{Finn, M.~D.}). In classical fluids, finite-order\index[subject]{finite-order} stirring protocols have been shown to stretch material lines linearly and therefore do not mix as efficiently as their pseudo-Anosov\index[subject]{pseudo-Anosov} counterparts.  

\begin{figure}[ht!]
\includegraphics[width=10cm]{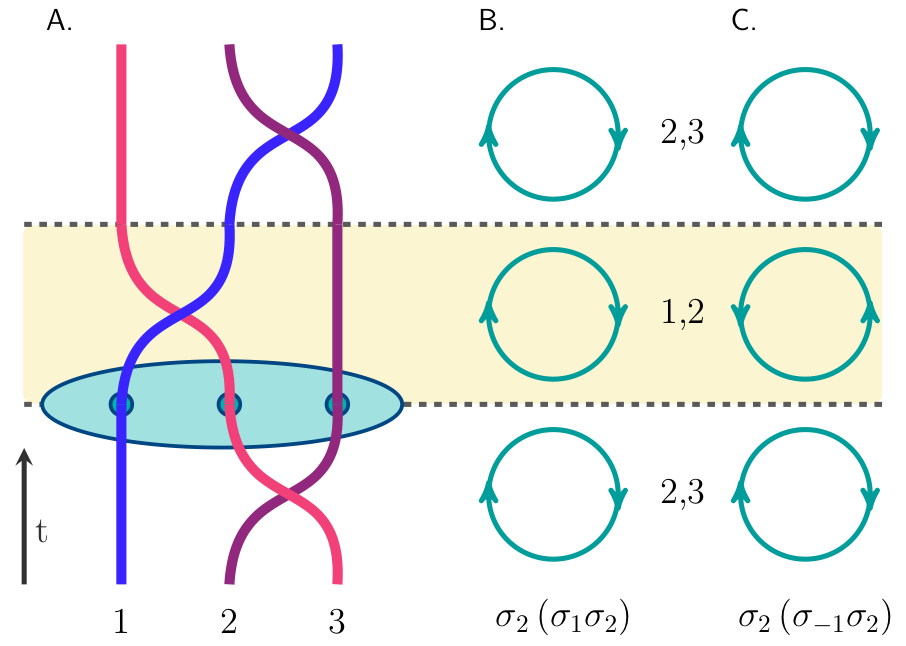}\hspace{0.2cm}
\begin{minipage}[b]{13pc}
\caption{\label{Braid} (A) Braid representation in space and time depicting the exchange of stirrers for a pseudo-Anosov\index[subject]{pseudo-Anosov} stirring protocol. Three stirring operations are shown, corresponding to the exchange of stirrers $\sigma_{2}=(2,3)$, then $\sigma_{-1}=(1,2)$ and again stirrers $\sigma_{2}=(2,3)$. The rotation stirrers trace out are depicted in (B) for a finite-order\index[subject]{finite-order} stirring protocol and (C) for the pseudo-Anosov\index[subject]{pseudo-Anosov} stirring protocol depicted in (A).}
\end{minipage}
\end{figure}

The trajectory of the stirring lasers can be represented by braid-diagrams in space-time, with braid letters $\sigma_{i}$ denoting the exchange of stirrers labelled $i$ and $i+1$ in a clockwise direction. This can be depicted in a braid diagram by the strand corresponding to the stirrer labelled $i+1$ passing over the strand showing the $i$th stirrer. Similarly, an anti-clockwise path of stirrers through the condensate is denoted by braid letters $\sigma_{-i}$ which represents stirrers $i+1$ and $i$ exchanging position in an anti-clockwise direction. This interchange is shown on a braid-diagram by the strand depicting the $i+1$ stirrer passing under the $i$th strand. Figure \ref{Braid} (A) illustrates the braid diagram corresponding to the exchange of stirrers following a protocol that can be classified as pseudo-Anosov\index[subject]{pseudo-Anosov}.  

Each stirring operation, corresponding to a braid letter, can be represented by a braid matrix. For the stirring operations corresponding to the $\sigma_{1}$, $\sigma_{-1}$ and $\sigma_{2}$ braid letters, which we apply in this work, the corresponding braid matrices take the form (\cite{Boyland2000,Finn2003}\index[authors]{Boyland, P.~L.}\index[authors]{Aref, H.}\index[authors]{Stremler, M.~A.}\index[authors]{Finn, M.~D.}\index[authors]{Cox, S.~M.}\index[authors]{Byrne, H.~M.})
\begin{equation}
{{\Sigma}}_{1}=\left[
\begin{array}{cc}
1 & -1 \\
0 & 1
\end{array} 
\right],\quad
{{\Sigma}}_{-1}=\left[
\begin{array}{cc}
1 & 1 \\
0 & 1
\end{array}
\right],\quad
{{\Sigma}}_{2}=\left[
\begin{array}{cc}
1 & 0 \\
1 & 1
\end{array}
\right].
\end{equation}
The time-like sequence of braid letters composes a braid word, where the rightmost letter gives the first stirring operation and the leftmost letter denotes the most recent stirring operation. We concentrate on two stirring protocols, which we label $A$ and $B$ that are represented by the braid words
\begin{equation}
A = \left(\sigma_{1}\sigma_{2}\right)^{n}\quad \text{and} \quad B = \left(\sigma_{-1}\sigma_{2}\right)^{n}\,, 
\end{equation}
here $n$ denotes the number of repetitions of braid letter operations. Similarly, the corresponding action of the braid matrices takes the form 
\begin{equation}
A = \left[\Sigma_{1}\Sigma_{2}\right]^{n} \quad \text{and} \quad B = \left[ \Sigma_{-1}\Sigma_{2}\right]^{n}\,.
\end{equation}
Stirring protocol $A$ is a finite-order\index[subject]{finite-order} protocol, while protocol $B$ is a pseudo-Anosov\index[subject]{pseudo-Anosov} stirring protocol which builds topological chaos\index[subject]{chaos} into the flow.  The  topology of the fluid flow is completely characterised by the braid word, regardless of the nature of the fluid flow, i.e. if it is compressible, incompressible, viscous or inviscid.  For three stirrers, the magnitude of the largest eigenvalue of the associated braid matrix,  called the spectral radius, gives a rate at which at least one material line is stretched.   \cite{Finn2003}\index[authors]{Finn, M.~D.}\index[authors]{Cox, S.~M.}\index[authors]{Byrne, H.~M.} computed the stretching in time of a finite material line element and showed that the chaotic region of the fluid flow is of the same extent as the region of fluid the stirrers are moved through. 

\begin{figure}[ht!]
\includegraphics[width=10cm]{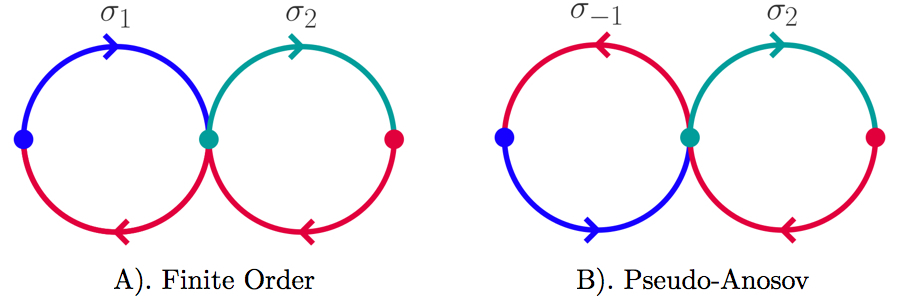}\hspace{0.2cm}
\begin{minipage}[b]{13pc}\vspace{-1cm}
\caption{\label{eight} Stirrer paths for first two stirring operations of A) finite-order\index[subject]{finite-order} stirring protocol and B) pseudo-Anosov\index[subject]{pseudo-Anosov} stirring protocol. }
\end{minipage}
\end{figure}

In the remainder of this paper we present numerical simulations of three laser stirrers in an atomic BEC\index[subject]{Bose-Einstein condensate} tracing out the finite-order\index[subject]{finite-order} and pseudo-Anosov\index[subject]{pseudo-Anosov} stirring schemes $A$ and $B$, presented above. Note the stirrers trace out a figure eight path, depicted in figure \ref{eight}. We choose the velocity of the laser stirrers to be larger than the critical velocity for vortex nucleation in order to determine if a pseudo-Anosov\index[subject]{pseudo-Anosov} stirring protocol ($B$) also results in a more random distribution of vortices\index[subject]{vortices} than a finite-order\index[subject]{finite-order} stirring scheme ($A$). 

\section{Modelling vortex dynamics}

\begin{figure}[h]
\includegraphics[width=\linewidth]{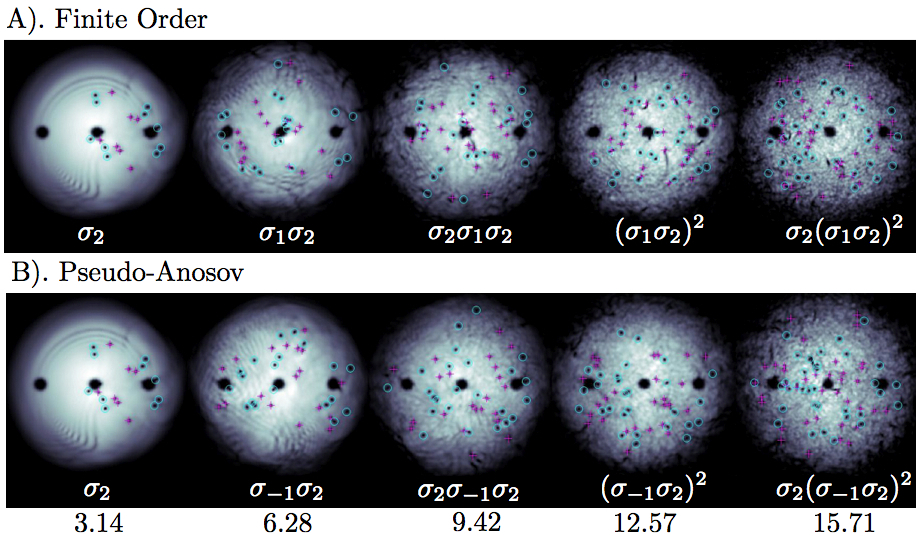}
\caption{\label{vortexdynamics}Vortex Dynamics: Condensate density profiles after simulation times given along the bottom of the figure. Magenta $+$ and cyan $-$ symbols denote vortices\index[subject]{vortices} with positive and negative winding respectively.  Row A.) finite-order\index[subject]{finite-order} stirring protocol at times corresponding to braid-words $\sigma_{2}$, $\sigma_{1}\sigma_{2}$, $\sigma_{2}\sigma_{1}\sigma_{2}$, $\left(\sigma_{1}\sigma_{2}\right)^{2}$ and $\sigma_{2}\left(\sigma_{1}\sigma_{2}\right)^{2}$.
Row B.) corresponds to pseudo-Anosov\index[subject]{pseudo-Anosov} stirring protocols at identical times, corresponding to braid words $\sigma_{2}$, $\sigma_{-1}\sigma_{2}$, $\sigma_{2}\sigma_{-1}\sigma_{2}$, $\left(\sigma_{-1}\sigma_{2}\right)^{2}$ and $\sigma_{2}\left(\sigma_{-1}\sigma_{2}\right)^{2}$.
}
\end{figure}

The second quantised hamiltonian describing a trapped, weakly-interacting atomic Bose-Einstein condensate\index[subject]{Bose-Einstein condensate} takes the form
\begin{equation}
H = \int d{\bf{\tilde{r}}}\left[
-\hat{\Psi}^{\dagger}\left({\bf{\tilde{r}}}\right)\frac{\hbar^{2}}{2m}\nabla^{2}\hat{\Psi}\left({\bf{\tilde{r}}}\right)
+\left(V\left({\bf{\tilde{r}}}\right)+V_{P}\right)\hat{\Psi}^{\dagger}\left({\bf{\tilde{r}}}\right)\hat{\Psi}\left({\bf{\tilde{r}}}\right)+\frac{U_{0}}{2}\hat{\Psi}^{\dagger}\left({\bf{\tilde{r}}}\right)\hat{\Psi}^{\dagger}\left({\bf{\tilde{r}}}\right)\hat{\Psi}\left({\bf{\tilde{r}}}\right)\hat{\Psi}\left({\bf{\tilde{r}}}\right)
\right] \,.
\end{equation}
Here $\hat{\Psi}\left(\bf{\tilde{r}}\right)^{\dagger}$ and $\hat{\Psi}\left(\bf{\tilde{r}}\right)$ are the creation and annihilation operators for bosons of mass $m$, trapped in a potential $V(\bf{\tilde{r}})$.  The laser stirrers are described by $V_{P}$.  The boson-boson interaction is given by a contact potential $U_{0}=4\pi\hbar^{2}a/m$,  where $a$ is the s-wave scattering length. The evolution of the field operator can be obtained from the Heisenberg equations of motion.  At low temperatures, we can describe the condensed state by a classical mean field $\Psi=\langle\hat{\Psi}\rangle$ and excitations $\langle\phi\rangle$, decomposing the boson operators as $\hat{\Psi}=\Psi+\langle\phi\rangle$. The dynamics of the mean-field condensate wavefunction are then given by the nonlinear Schr\"{o}dinger equation
 \begin{equation}
i \hbar\frac{\partial \Psi\left(\bf{\tilde{r}}\right)}{\partial \tilde{t}} = -\frac{\hbar^{2}}{2m}\nabla^{2}\Psi\left(\bf{\tilde{r}}\right)+\left(V\left(\bf{\tilde{r}}\right)+V_{P}\right)\Psi\left(\bf{\tilde{r}}\right) + U_{0} |\Psi\left(\bf{\tilde{r}}\right)|^{2}\Psi\left(\bf{\tilde{r}}\right)\,.
 \end{equation}
We consider a condensate trapped such that the axial trapping frequency is much greater than the radial trapping frequency, $ \omega_{z}\gg\omega_{r}$. In this case the condensate is quasi-two dimensional.  For our simulations we model the dimensionless form of the 2D non-linear Schr{\"o}dinger equation, 
  \begin{equation}\label{gpe}
i \frac{\partial \psi}{\partial t} = -\frac{1}{2}\nabla^{2}\psi + \frac{{r}^{2}}{2}\psi+V_{P}\psi + \kappa_{2d} |\psi|^{2}\psi\,.
\end{equation}
Here the 2D interaction strength takes the form $\kappa_{2d}= 2\sqrt{2\pi}aN / a_{z} $, where $N$ is the total number of atoms in the condensate. The 2D condensate wavefunction is written as $\psi=a_{r}\Psi/\sqrt{N}$ and we have imposed the normalisation $\int\text{d\bf{x}}|\psi|^{2}=1$. The radial (axial) trapping frequency, $\omega_{r}$ ($\omega_{z}$), determines the radial (axial) harmonic oscillator length  $a_{r}= \sqrt{\hbar/(m\omega_{r})}$ ($a_{z}= \sqrt{\hbar/(m\omega_{z})}$). We choose the scaling for lengths and time to be  $\tilde{r}/a_{r}=r$ and $\tilde{t}\omega_{r}=t$, where $\tilde{r}$ and $\tilde{t}$ are dimensional with units of meters and seconds respectively.  We solve Eq.({\ref{gpe}}) pseudo-spectrally, applying an adaptive 4-5th order Runge-Kutta method in time (\cite{Dennis2012}). We select a stirring strength $V_{0}\sim2.6\mu$, where $\mu$ is the chemical potential of the BEC. We choose $\kappa=10400$, which taking $^{23}$Na atoms corresponds to parameters of $\omega_{z}=2\pi\times50$Hz, $\omega_{r}=2\pi\times5$Hz, $N=2.2\times10^{6}$ and $a=2.75$nm. Simulations are run on a $512^{2}$ grid of spatial extent $-20$ to $20$.  Snap-shots of time-evolution of the condensate density profile is depicted in figure \ref{vortexdynamics}.

\subsection{Decomposition of the Kinetic Energy contribution}

 For a time-independent potential, the total energy of the system is conserved within the nonlinear Schr{\"o}dinger equation. Writing the mean-field condensate wavefunction in terms of the condensate density $n=|\psi|^{2}$, and phase $\theta$, as $\psi=\sqrt{n}\exp\left[i\theta\right]$, the kinetic energy component of the total energy can be decomposed into contributions from the  quantum pressure and kinetic energy density (\cite{Nore1997}\index[authors]{Nore, C.}\index[authors]{Abid, M.}\index[authors]{Brachet, M.~E.})
 \begin{equation}
 E_{\sf KE}=\int \text{d\bf{x}}\left( \frac{1}{2}|\nabla\psi|^{2}\right)=\int \text{d\bf{x}}\left( \frac{1}{2}|\nabla\sqrt{n}|^{2}+\frac{1}{2}|\sqrt{n}\textbf{v}|^{2}\right)\,.
 \end{equation}
 Here we have defined the velocity field $\textbf{v}=\nabla\theta$. The kinetic energy density contribution can be further decomposed into a compressible component for which the density weighted velocity field ${\bf{\Upsilon}}=\sqrt{n}\textbf{v}$, satisfies $\nabla\times{\bf{\Upsilon}}^{c}=0$ and an incompressible, divergence free component $\nabla\cdot{\bf{\Upsilon}}^{i}=0$, such that ${\bf{\Upsilon}}={\bf{\Upsilon}}^{i}+{\bf{\Upsilon}}^{c}$.  The total compressible energy is a useful measure of the total sound in the system and the total incompressible kinetic energy gives the total kinetic energy of vortices\index[subject]{vortices}  
 \begin{equation}
 E^{i,c}_{\sf KE}=\frac{1}{2}\int\text{d}\textbf{x}\left|{\bf{\Upsilon}}^{i,c}\right|^{2} \,.
 \end{equation}
 
 One can subsequently define compressible and incompressible kinetic energy spectrums, however due to the small size of trapped atomic condensates, only a short range of length scales is accessible in these systems, in comparison to other 2D turbulent systems.  Consequently, evaluating scaling exponents from such spectrums is of limited use and the derived scaling laws should be interpreted with care. We refrain from such an analysis here. 

\begin{figure}[!ht]
\begin{minipage}{0.6\linewidth}
\includegraphics[width=\linewidth]{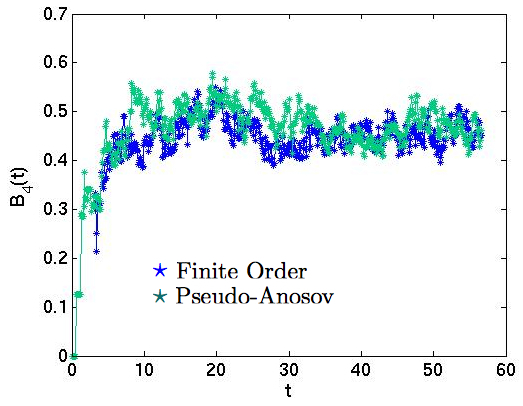}
\end{minipage}
\begin{minipage}{0.3\linewidth}
\caption{\label{clust} Evolution of vortex distribution statistics given by fourth nearest neighbor vortices for finite-order (blue $\ast$) and pseudo-Anosov (green $\ast$) stirring schemes.}
\end{minipage}
\end{figure}
 
\subsection{Statistical measures of clustering\index[subject]{clustering}} 
 
Motivated by the seminal work of \cite{Ripley1976,Ripley1977}\index[authors]{Ripley, B.~D.}\index[authors]{Ripley, B.~D.}, \cite{White2012}\index[authors]{White, A.~C.}\index[authors]{Barenghi, C.~F.}\index[authors]{Proukakis, N.~P.} developed statistical measures of clustering\index[subject]{clustering} that are appropriate for measuring if vortices\index[subject]{vortices} are clustered or more randomly distributed in atomic Bose-Einstein condensates\index[subject]{Bose-Einstein condensate}. These measures have the advantage of being experimentally accessible as they rely only on knowledge of the position and winding of vortices\index[subject]{vortices}, taking into account both the sign and distance between a reference vortex $i$ and its neighbouring vortices\index[subject]{vortices} $j$, up to its $J^{\text{th}}$ nearest-neighbor,  
\begin{equation}\label{statclust}
B_{J}(t)=\frac{1}{N}\sum_{i=1}^{N}\sum_{j=1}^{J}\frac{b_{ij}(t)}{J}\,.
\end{equation} 
$N$ denotes all vortices\index[subject]{vortices} that are at a distance $d_{i} \geq R_{E}$ from the condensate edge.  If vortex $i$ and its $j^{th}$ nearest neighbor are of opposite sign, $b_{ij}=0$.  If vortex $i$ and its $j^{th}$ nearest neighbor are of the same sign and separation distance between them, $d_{ij}\leq R_{c}$, then $b_{ij}=1$. If vortex $i$ and vortex $j$ are separated by a distance larger than $R_{c}$,  ($d_{ij}\geq R_{c}$), then $b_{ij}=0$.  In this work we take the distance $R_{c}=R_{E}$ to be greater than the average inter-vortex separation distance and of the order of the largest cluster size.  These constraints are included in the statistical measure of vortex distribution in Eq.(\ref{statclust}), to account for edge effects that arise as we are dealing with a small system of finite size. The parameter limiting separation distance between vortices\index[subject]{vortices} $i$ and $j$ gives a means to investigate the variation of clustering\index[subject]{clustering} over spatial regions. We take vortex distribution to be clustered for values of  $B_J(t)>0.5$, while for more random vortex distributions $B_J(t)<0.5$. Related measures of clustering\index[subject]{clustering} have been recently applied to look at two dimensional turbulence\index[subject]{turbulence} in \cite{Bradley2012}\index[authors]{Bradley, A.~S.}\index[authors]{Anderson, B.~P.} and \cite{Reeves2012arxiv}\index[authors]{Reeves, M.~T.}\index[authors]{Billam, T.~B.}\index[authors]{Anderson, B.~P.}\index[authors]{Bradley, A.~S.}. 

\begin{figure}[!ht]
\begin{minipage}{0.6\linewidth}
\includegraphics[width=\linewidth]{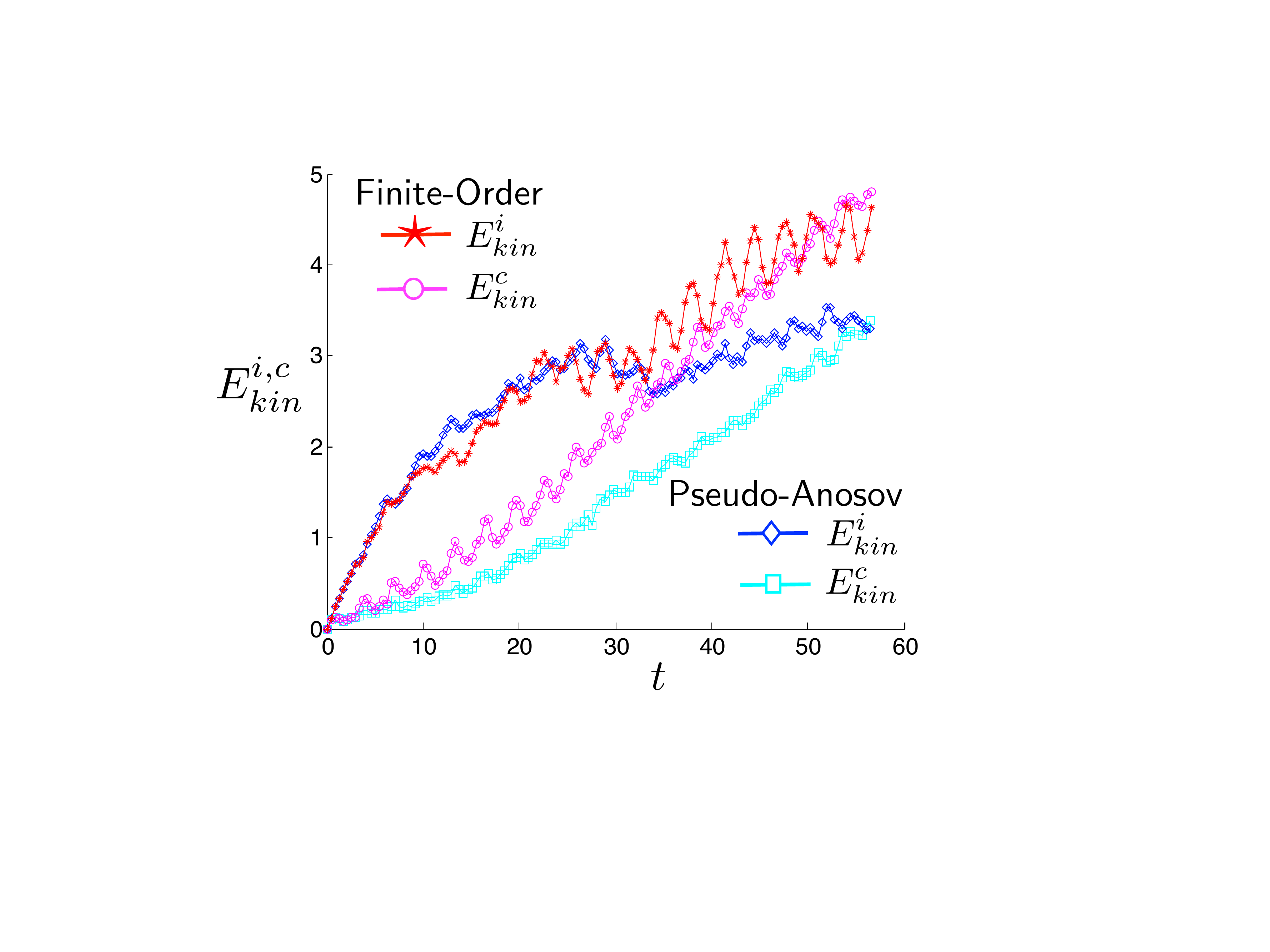}
\end{minipage}
\begin{minipage}{0.3\linewidth}
\caption{\label{kic} Incompressible and compressible kinetic energy distribution for finite-order stirring given by red stars and magenta circles respectively, and pseudo-Anosov stirring, given by  blue diamonds and cyan squares respectively.}
\end{minipage}
\end{figure}

\section{Discussion and Conclusions}

From the point of view of classical fluid dynamics, visualising and tracking the evolution of fluid material lines through the use of tracer particles such as dye, are standard and straightforward techniques that enable one to distinguish between underlying chaotic\index[subject]{chaos} and regular dynamics in the fluid flow. For atomic Bose-Einstein condensates\index[subject]{Bose-Einstein condensate}, the best method for flow visualisation is not so obvious, however some information about the condensate flow can be provided with current experimental techniques.  The positions of vortices\index[subject]{vortices} can be visualised through standard absorption imaging techniques and the condensate phase profile has the potential to be measured by applying interferometric measurements giving information about vortex\index[subject]{vortices} winding. For these reasons we employ the statistical measure of clustering\index[subject]{clustering} introduced above to analyse condensate flow. 

We employ Eq.({\ref{statclust}}) up to fourth-nearest neighbor vortices\index[subject]{vortices} and track the corresponding statistical measure of clustering\index[subject]{clustering}, which is plotted in figure {\ref{clust}}. Evidently, neither the finite-order\index[subject]{finite-order} nor the pseudo-Anosov\index[subject]{pseudo-Anosov} stirring protocol that we have chosen results in a vortex distribution that is significantly clustered. After a few stirring operations, the   vortices\index[subject]{vortices} appear randomly distributed, independent of stirring protocol.  This result is an indication of the potential nature of quantum fluids. One might expect underlying chaotic dynamics would result in a vortex distribution that is less clustered than underlying regular dynamics. Although material line elements are expected to be exponentially stretched for pseudo-Anosov\index[subject]{pseudo-Anosov} stirring in contrast to the linear stretching of material line elements in the finite-order\index[subject]{finite-order} case, this does not appear to influence the vortex distribution. In other words, the vortex distribution is independent of if topological chaos\index[subject]{chaos} is built into the flow by the method of stirring. 

Another explanation for our observations may also lie in the speed at which the condensate is stirred. From a classical fluid perspective, when a container of classical fluid, such as water is stirred with very fast moving obstacles, the fluid will become turbulent\index[subject]{turbulence} and well-mixed, independent of the motion of the stirring obstacles provided their paths traverse a significant extent of the container surface. In this limit linear or exponential stretching of material line elements in the fluid may be present, but not visible. For a quantum fluid we can define limits on the speed at which the condensate is stirred. The lower limit on stirring speed is given by the condensate critical velocity, below which an obstacle moving through the condensate will not nucleate vortices. The upper limit is set by the speed at which fragmentation of the condensate occurs (\cite{Parker2005}\index[authors]{Parker,N~G.}\index[authors]{Adams, C.~S.}). In our simulations the condensate is stirred at speeds faster than the critical velocity for vortex\index[subject]{vortices} nucleation, but not so fast that we enter the regime of condensate fragmentation. Although the speed of stirring is chosen to be fast enough to nucleate vortices\index[subject]{vortices}, this does not necessarily imply we are in the quantum analogue of the classical scenario described above, where linear or exponential stretching of material lines may be present but not visible due to the very fast stirring speeds. Defining the analogue of such a limit is an interesting question we leave for future work. 

\begin{figure}[!ht]
\begin{minipage}{0.6\linewidth}
\includegraphics[width=\linewidth]{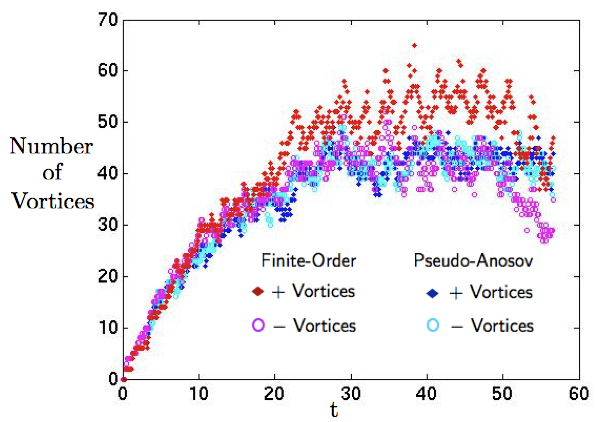}
\end{minipage}
\begin{minipage}{0.3\linewidth}
\caption{\label{vortnumber} Evolution of number of vortices with positive and negative winding given by red diamonds and magenta circles respectively for finite-order stirring and blue diamonds and blue circles respectively for pseudo-Anosov stirring.}
\end{minipage}
\end{figure}

From the analysis of the incompressible and compressible kinetic energy distribution (see figure \ref{kic}), it is found that pseudo-Anosov\index[subject]{pseudo-Anosov} stirring results in less sound being generated in comparison to finite-order\index[subject]{finite-order} stirring protocols.  The finite-order\index[subject]{finite-order} stirring scheme creates more sound in the condensate as result of the abrupt change in stirring direction during sequential stirring operations in this scheme.  Although stirrers in both schemes follow trajectories that trace out a figure of eight shape,  such abrupt changes in stirring direction do not occur in the pseudo-Anosov\index[subject]{pseudo-Anosov} stirring scheme employed.  Although at times $t<30$ the total incompressible kinetic energy is initially similar for both stirring protocols, at later times more incompressible kinetic energy is generated for finite-order\index[subject]{finite-order} stirring. This difference is also reflected in the total vortex number for both stirring protocols (refer to figure \ref{vortnumber}). Finite-order\index[subject]{finite-order} stirring generates a larger imbalance in the total number of positive and negative vortices\index[subject]{vortices} than pseudo-Anosov\index[subject]{pseudo-Anosov} stirring. The oscillations in the vortex number and kinetic energy distribution for finite order stirring reflect the oscillations in the total condensate volume induced by this stirring protocol.  

In conclusion, we have applied both pseudo-Anosov\index[subject]{pseudo-Anosov} and finite-order\index[subject]{finite-order} stirring protocols to introduce vortices\index[subject]{vortices} in a quasi-two dimensional Bose-Einstein condensate\index[subject]{Bose-Einstein condensate}.  Efficient mixing, as described by exponential stretching of material lines, could be thought to also translate into efficient vortex mixing. However, we observe that for a trapped inhomogeneous Bose-Einstein condensate\index[subject]{Bose-Einstein condensate}, there is no appreciable increase in `mixing' of vortices\index[subject]{vortices}, as determined by the random, or clustered nature of the vortex distribution.  That is, we do not find any significant correlation between if topological chaos is being introduced into the flow and how random the resulting vortex distribution is.  In future work, we set up a stirring scheme that induces topological chaos\index[subject]{chaos} in three dimensional quantum fluid flow and also analyse the onset of chaos\index[subject]{chaos} for both finite-order\index[subject]{finite-order} and pseudo-Anosov\index[subject]{pseudo-Anosov} stirring schemes.  

\ack
We thank the anonymous reviewer for their insightful comments and especially for raising the alternative explanation for our observations we have discussed.  This work was supported by the EPSRC. ACW acknowledges funding from EPSRC grant No.~EP/H027777/1 and  CFB and NPP acknowledge funding from EPSRC grant No.~EP/I019413/1.

\bibliographystyle{jfm2}

\bibliography{TOD-bibliography-White_AC}
\printindex[subject]
\printindex[authors]
\end{document}